\documentclass[fleqn,10pt]{wlscirep}
\usepackage[utf8]{inputenc}
\usepackage[T1]{fontenc}
\usepackage{color}    
\usepackage{graphicx}
\usepackage{dcolumn}
\usepackage{bm}
\usepackage{subfigure}
\usepackage{amssymb}
\usepackage{multirow}
\usepackage{amsmath}
\graphicspath{{plots/}}
\usepackage{hyperref}
\hypersetup{
    colorlinks = true,
    citecolor = blue
}
\usepackage[utf8]{inputenc}
\usepackage{wrapfig}
\usepackage[T1]{fontenc}
\usepackage[normalem]{ulem}
	
\usepackage{wrapfig}

\title{
X-ray Thomson scattering absolute intensity from the f-sum rule in the
imaginary-time domain
}

\author[1,2,*]{T.~Dornheim}
\author[3]{T.~D\"oppner}
\author[4]{A.~D.~Baczewski}
\author[5]{P.~Tolias}
\author[1,2,6]{M.~P.~B\"ohme}
\author[1,2]{Zh.~A.~Moldabekov}
\author[1,2]{Th.~Gawne}
\author[7]{D.~Ranjan}
\author[8]{D.~A.~Chapman}
\author[3]{M.~J.~MacDonald}
\author[9]{Th.~R.~Preston}
\author[7]{D.~Kraus}
\author[2]{J.~Vorberger}

\affil[1]{Center for Advanced Systems Understanding (CASUS), D-02826 G\"orlitz, Germany}
\affil[2]{Helmholtz-Zentrum Dresden-Rossendorf (HZDR), D-01328 Dresden, Germany}
\affil[3]{Lawrence Livermore National Laboratory (LLNL), California 94550 Livermore, USA}
\affil[4]{Center for Computing Research, Sandia National Laboratories, Albuquerque NM 87185, USA}
\affil[5]{Space and Plasma Physics, Royal Institute of Technology (KTH), Stockholm SE-100 44, Sweden}
\affil[6]{Technische Universit\"at Dresden, D-01062 Dresden, Germany}
\affil[7]{Institut f\"ur Physik, Universit\"at Rostock, D-18057 Rostock, Germany}
\affil[8]{First Light Fusion, Yarnton, Oxfordshire, United Kingdom}
\affil[9]{European XFEL, D-22869 Schenefeld, Germany}
\affil[*]{t.dornheim@hzdr.de}

\begin{abstract}We present a formally exact and simulation-free approach for the normalization of X-ray Thomson scattering (XRTS) spectra based on the f-sum rule of the imaginary-time correlation function (ITCF). Our method works for any degree of collectivity, over a broad range of temperatures, and is applicable even in nonequilibrium situations. In addition to giving us model-free access to electronic correlations, this new approach opens up the intriguing possibility to extract a plethora of physical properties from the ITCF based on XRTS experiments.
\end{abstract}
\begin{document}

\flushbottom
\maketitle
%
%
\thispagestyle{empty}

\section*{Introduction}

Matter at extreme temperatures ($T\sim10^{3}-10^8\,$K) and pressures ($P\sim1-10^{4}\,$Mbar) is ubiquitous throughout nature~\cite{drake2018high,fortov_review} and occurs within astrophysical objects such as giant planet interiors~\cite{Benuzzi_Mounaix_2014} and brown dwarfs~\cite{becker}. 
In addition, such high energy density (HED) conditions can be realized in the laboratory using different techniques~\cite{falk_wdm,nguyen2022direct}. Prominent examples include inertial confinement fusion~\cite{hu_ICF} as it is realized at the National Ignition Facility (NIF)~\cite{Moses_NIF}, and isochoric heating using free-electron X-ray laser beams~\cite{kraus_xrts}. 
Indeed, there have been a number of spectacular experimental achievements over the last years, including the observation of diamond formation at planetary interior conditions by Kraus~\emph{et al.}~\cite{Kraus2016,Kraus2017} and a number of breakthroughs related to nuclear fusion at NIF~\cite{PhysRevLett.129.075001,Zylstra2022,icf-collab_prl_24}.

However, the extreme conditions render the diagnostics a formidable challenge as often even basic parameters such as the temperature cannot be directly measured. Instead, they have to be inferred indirectly from other observations. In this regard, the X-ray Thomson scattering (XRTS) technique~\cite{sheffield2010plasma,siegfried_review} has emerged as a widely used tool that is available both in the HED regime~\cite{kraus_xrts,Glenzer_PRL_2007,DOPPNER2009182} and at ambient conditions~\cite{Abela:77248}.
The measured XRTS intensity $I(\mathbf{q},\omega)$ (with $\mathbf{q}$ and $\omega$ being the scattering momentum and energy, respectively) can be expressed as~\cite{Dornheim_T2_2022}
\begin{eqnarray}\label{eq:convolution}
I(\mathbf{q},\omega) =A\, S_{ee}(\mathbf{q},\omega) \circledast R(\omega),
\end{eqnarray}
with $S_{ee}(\mathbf{q},\omega)$ being the electron--electron dynamic structure factor and $R(\omega)$ being the combined source and instrument function (SIF). 
The latter is often known from either separate source monitoring or the characterization of backlighter sources~\cite{MacDonald_POP_2022}. Moreover, $A$ denotes a normalization constant that is \emph{a priori} unknown.

In practice, the numerical deconvolution of Eq.~(\ref{eq:convolution}) to extract $S_{ee}(\mathbf{q},\omega)$ is unstable due to the noise in the experimental data. Therefore, the traditional way to interpret an XRTS signal is to construct a model $S_\textnormal{model}(\mathbf{q},\omega)$ and then convolve it with $R(\omega)$. Comparing with the experimental signal then allows one to obtain the unknown variables such as the temperature, which are being treated as free parameters~\cite{Kasim_POP_2019}. On the one hand, this \emph{forward modeling} procedure gives one access e.g.~to the equation-of-state~\cite{Falk_HEDP_2012,Falk_PRL_2014}, which is of prime importance for a gamut of practical applications. On the other hand, the inferred system parameters can strongly depend on the employed model assumptions, such as the Chihara decomposition into \emph{bound} and \emph{free} electrons~\cite{Chihara_1987,Gregori_PRE_2003}.

Very recently, Dornheim \emph{et al.}~\cite{Dornheim_T_2022,Dornheim_T2_2022,Dornheim_insight_2022} have suggested to instead analyze the two-sided Laplace transform of $S_{ee}(\mathbf{q},\omega)$,
\begin{eqnarray}\label{eq:Laplace}
F_{ee}(\mathbf{q},\tau) = \mathcal{L}\left[S_{ee}(\mathbf{q},\omega)\right] =\int_{-\infty}^\infty \textnormal{d}\omega\ e^{-\hbar\omega\tau} S_{ee}(\mathbf{q},\omega)\ ,
\end{eqnarray}
which has a number of key advantages. First, the deconvolution with respect to the SIF is straightforward in the Laplace domain, see Eq.~(\ref{eq:convolution_theorem}) below. Second, the Laplace transform is remarkably robust with respect to noise in the intensity~\cite{Dornheim_T2_2022}. Third, Eq.~(\ref{eq:Laplace}) directly corresponds to the imaginary-time correlation function (ITCF) $F_{ee}(\mathbf{q},\tau)$~\cite{Dornheim_insight_2022}. The latter corresponds to the usual intermediate scattering function $F_{ee}(\mathbf{q},t)$, but evaluated at an imaginary argument $t=-i\hbar\tau$, with $\tau\in[0,\beta]$ and $\beta=1/k_\textnormal{B}T$ being the usual inverse temperature.
In thermodynamic equilibrium, $F_{ee}(\mathbf{q},\tau)$ is symmetric around $\tau=\beta/2$, which means that Eq.~(\ref{eq:Laplace}) allows one to extract the temperature without any models or approximations~\cite{Dornheim_T_2022,Dornheim_T2_2022}.


While the ITCF, by definition, contains exactly the same information as $S_{ee}(\mathbf{q},\omega)$~\cite{Dornheim_insight_2022,Dornheim_review}, knowing its proper normalization is crucial for the extraction of other system parameters beyond the temperature. For example, the relation $S_{ee}(\mathbf{q})=F_{ee}(\mathbf{q},\tau=0)$ then gives one direct access to the electron--electron static structure factor. Moreover, absolute knowledge of the ITCF 
will potentially open the way toward experimental measurements of the exchange--correlation kernel [cf.~Eq.~(\ref{eq:chi})] of real materials, which will be of high value for the benchmarking and further development of \emph{ab initio} density functional theory (DFT) simulations~\cite{Moldabekov_JCTC_2023, Moldabekov_non_empirical_hybrid, Maximilian_2023, JCP_hybrids_2023}.
Other important parameters that are encoded into $F_{ee}(\mathbf{q},\tau)$, but cannot be obtained from its symmetry alone, include the number density $n$ and the ionization degree $Z$, which are of key importance for equation-of-state tables~\cite{Falk_PRL_2014,Falk_HEDP_2012}.

In this work, we overcome this obstacle without the need for empirical parameters, models or approximations.
Specifically, we demonstrate how to determine the normalization of an XRTS signal by utilizing the well-known f-sum rule in the imaginary-time domain. Our method works for all wave numbers covering both the collective and the single-particle regime. We apply it to measurements at HED conditions taken at NIF~\cite{Tilo_Nature_2023} and ambient conditions investigated at the European XFEL~\cite{Voigt_POP_2021}. Finally, the general validity of the f-sum rule makes it available even in the case of nonequilibrium~\cite{Vorberger_PRX_2023}.

\section*{Theory}

\begin{figure*}\centering
\includegraphics[width=0.462\textwidth]{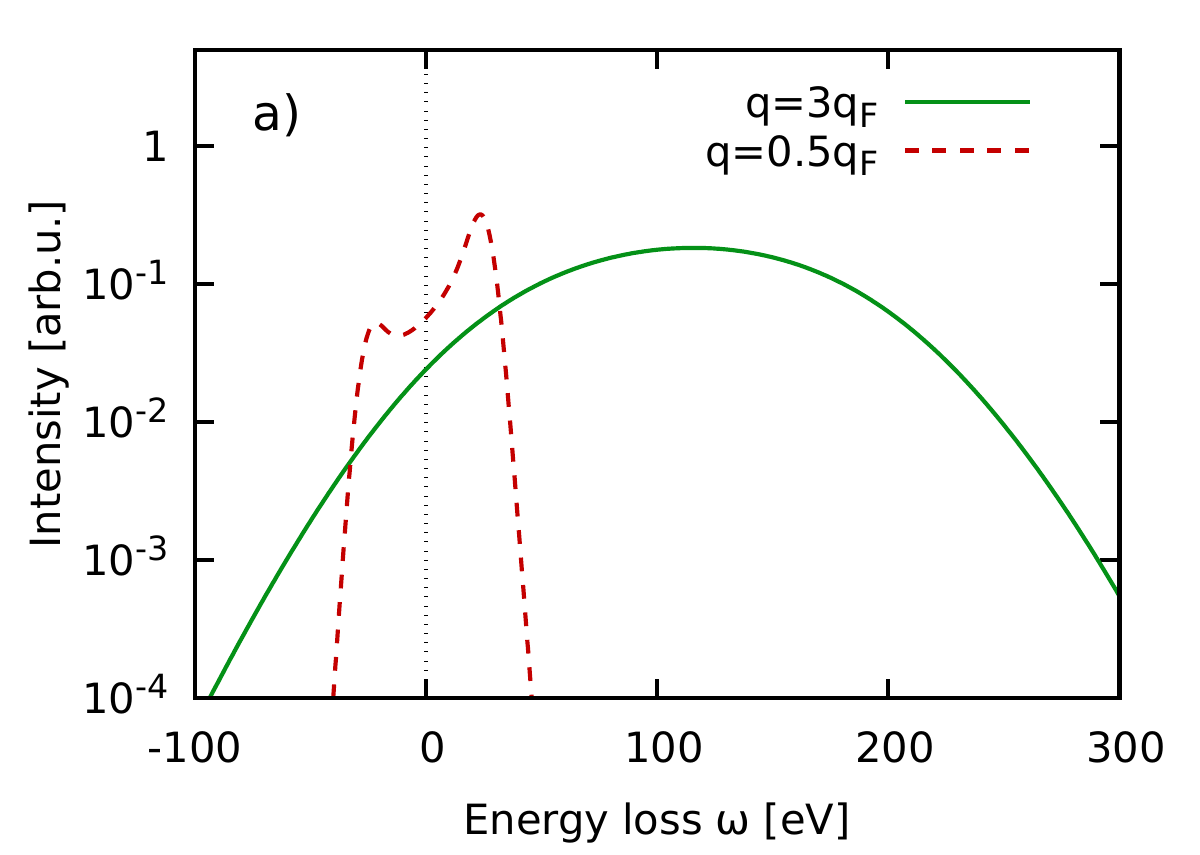}\includegraphics[width=0.462\textwidth]{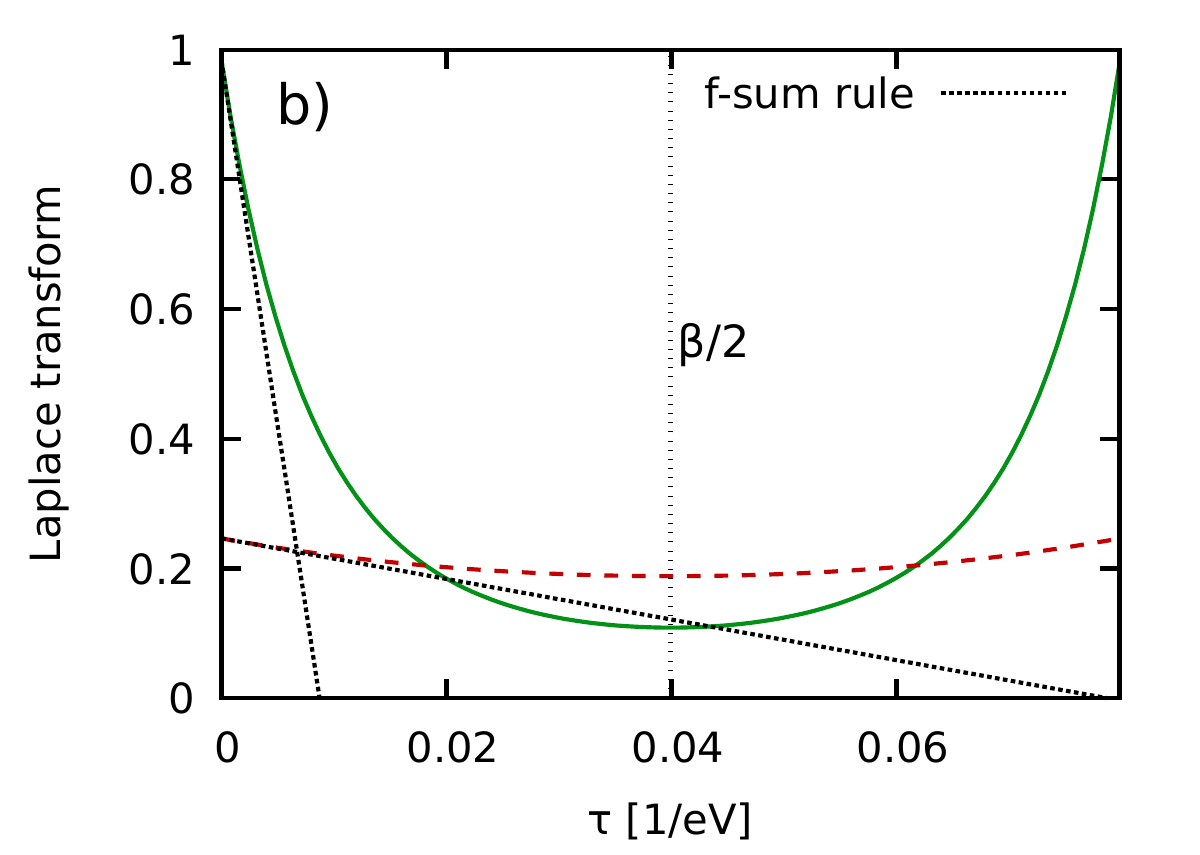}
\caption{\label{fig:UEG}
 a) Synthetic XRTS intensity for a uniform (free) electron gas at $r_s=2$ and $\Theta=1$ ($T=12.53\,$eV) with the Fermi wave number $q_\textnormal{F}=1.81\,$\AA$^{-1}$. The dynamic structure factors $S_{ee}(\mathbf{q},\omega)$ for $q=3q_\textnormal{F}$ (solid green) and $q=0.5q_\textnormal{F}$ (dashed red) have been convolved with a Gaussian instrument function of width $\sigma=3.33\,$eV. b) Corresponding ITCF, cf.~Eq.~(\ref{eq:Laplace}). The dotted black lines show asymptotic linear expansions around $\tau=0$ and have been obtained from the f-sum rule.
}
\end{figure*}

A key advantage of the two-sided Laplace transform defined in Eq.~(\ref{eq:Laplace}) above is the convolution theorem
\begin{eqnarray}\label{eq:convolution_theorem}
AF_{ee}(\mathbf{q},\tau)=A\mathcal{L}\left[S_{ee}(\mathbf{q},\omega)\right] = \frac{\mathcal{L}\left[I(\mathbf{q},\omega) \right]}{\mathcal{L}\left[R(\omega)\right]} \,,
\end{eqnarray}
which gives us straightforward access to $A F_{ee}(\mathbf{q},\tau)$. The numerical stability of Eq.~(\ref{eq:convolution_theorem}) with respect to experimental noise has been analyzed in detail in Ref.~\cite{Dornheim_T2_2022}.

To determine the normalization constant $A$, we consider the frequency moments of the dynamic structure factor~\cite{Dornheim_review}
\begin{eqnarray}\label{eq:moments}
M_\alpha^{S}  = \int_{-\infty}^\infty \textnormal{d}\omega\ S_{ee}(\mathbf{q},\omega)\ \omega^\alpha\ .
\end{eqnarray}
It is easy to see that all positive integer $M_\alpha^{S}$ can be obtained from $\tau$-derivatives of the ITCF around $\tau=0$~\cite{Dornheim_insight_2022,Dornheim_moments_2023},
\begin{eqnarray}\label{eq:moments_derivative}
M_\alpha^{S}= \frac{\left( -1 \right)^\alpha}{\hbar^\alpha}  \left. \frac{\partial^\alpha}{\partial\tau^\alpha} F_{ee}(\mathbf{q},\tau) \right|_{\tau=0} \ .
\end{eqnarray}
We note that Eq.~(\ref{eq:moments_derivative}) has been employed in the recent Ref.~\cite{Dornheim_moments_2023} to extract $M^S_\alpha$ up to fifth order from \emph{ab initio} path integral Monte Carlo (PIMC) simulations of the warm dense electron gas over a broad range of wavenumbers.
The final ingredient to our approach is given by the well-known f-sum rule~\cite{quantum_theory} $M_1^S=\hbar q^2/2m_\textnormal{e}$, which, when being combined with Eqs.~(\ref{eq:convolution_theorem}) and (\ref{eq:moments_derivative}), leads to the relation
 \begin{eqnarray}\label{eq:final}
     A = \left. - \frac{2m_e}{(\hbar q)^2} \frac{\partial}{\partial\tau} \frac{\mathcal{L}\left[I(\mathbf{q},\omega)\right]}{\mathcal{L}\left[R(\omega)\right]}\right|_{\tau=0}\ .
 \end{eqnarray}
Eq.~(\ref{eq:final}) implies that we can directly calculate $A$ from the slope of the ratio of the Laplace transforms of the intensity and the SIF around the origin.

The physical manifestation of this idea is demonstrated in Fig.~\ref{fig:UEG} for the uniform electron gas at the metallic density of $r_s=2$ at the electronic Fermi temperature $\Theta=k_\textnormal{B}T/E_\textnormal{F}=1$, i.e., $T=12.53\,$eV. In panel a), we show synthetic intensities that have been obtained by convolving the dynamic structure factor of the UEG with a Gaussian model SIF of width $\sigma=3.33\,$eV. The solid green curve corresponds to the non-collective single-particle regime where the entire spectrum is described by a single inverse parabola, and the dashed red curve to half the Fermi wave number, which is mostly collective and results in a distinct peak around the plasma frequency. 
Panel b) shows the corresponding deconvolved ITCFs, which are symmetric around $\tau=\beta/2$; this symmetry relation directly follows from the well-known detailed balance relation $S_{ee}(\mathbf{q},-\omega)=S_{ee}(\mathbf{q},\omega)e^{-\beta\hbar\omega}$ that holds in thermal equilibrium. Indeed, this relation allows one, in principle, to extract the temperature from a measured XRTS intensity without the need for models or simulations~\cite{Dornheim_T_2022,Dornheim_T2_2022}.
In the present work, we exclusively focus on the vicinity of $\tau=0$,
where the dashed black lines depict the f-sum rule that describes the slope of $F_{ee}(\mathbf{q},\tau)$ around the origin. 
An imaginary-time diffusion perspective on the physical meaning of $F_{ee}(\mathbf{q},\tau)$ has been presented in the recent Ref.~\cite{Dornheim_PTR_2022}.

\begin{figure}\centering
\includegraphics[width=0.462\textwidth]{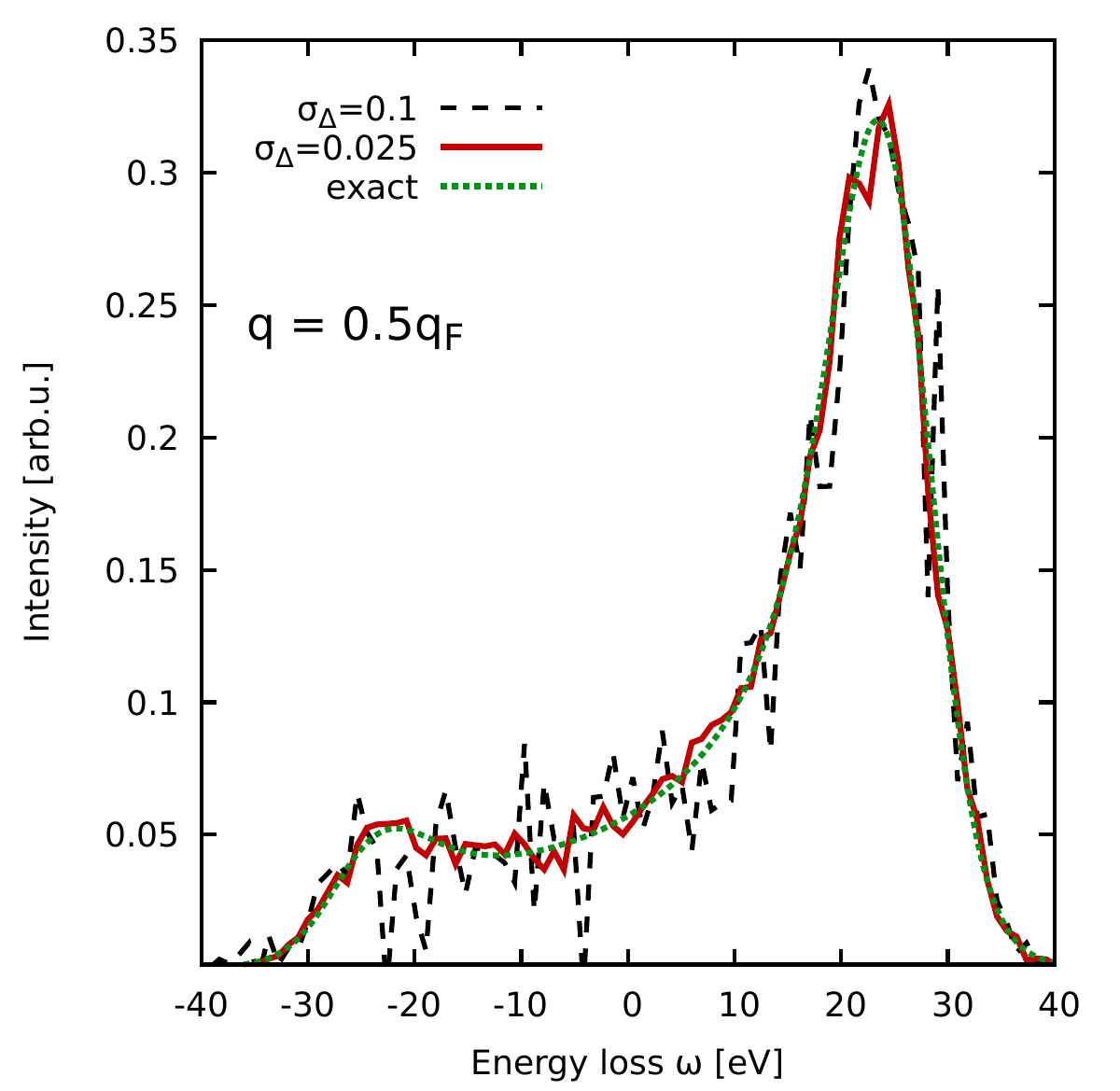}
\caption{\label{fig:synthetic_noise}
 Synthetic intensity computed for a pure UEG model~\cite{dornheim_dynamic,Dornheim_PRL_2020_ESA} with $r_s=2$ and $\Theta=1$ ($T=12.53\,$eV) for $q=0.5q_\textnormal{F}$. The dynamic structure factor has been convolved with a Gaussian instrument function of width $\sigma=3.33\,$eV. Dotted green: exact data; solid red and dashed black: perturbed with different noise levels $\sigma_\Delta$, cf.~Eq.~(\ref{eq:I_error}).
}
\end{figure}

While the basic idea to use the f-sum rule to normalize XRTS measurements has been reported in the literature~\cite{siegfried_review,GarciaSaiz2008}, the present approach combines two key advantages. First, it does not require an explicit deconvolution in the $\omega$-domain to get $S_{ee}(\mathbf{q},\omega)$, which would be required to directly evaluate Eq.~(\ref{eq:moments}). Second, it does also not presuppose any decomposition into effectively \emph{elastic} and \emph{inelastic} contributions to the signal as it was assumed e.g.~in Ref.~\cite{GarciaSaiz2008}.

\section*{Results}

\subsection*{Synthetic scattering data}\label{sec:synthetic}

An additional important question is the stability of the numerical derivative in Eq.~(6) in the main text with respect to experimental noise.
Following Refs.~\cite{sheffield2010plasma,Dornheim_T2_2022}, we express the experimental signal $I_\textnormal{exp}(\mathbf{q},\omega)$ as~\cite{sheffield2010plasma,Dornheim_T2_2022}\begin{eqnarray}\label{eq:I_error}
    I_\textnormal{exp}(\mathbf{q},\omega) = I(\mathbf{q},\omega) + \xi_{\sigma_\Delta}(\omega)\sqrt{I(\mathbf{q},\omega)}\,,
\end{eqnarray}
with $\xi_{\sigma_\Delta}(\omega)$ being a Gaussian random variable distributed around zero with $\sigma_\Delta$ being the variance.
In Fig.~\ref{fig:synthetic_noise}, we again consider the UEG model shown in Fig.~\ref{fig:UEG}, with the dotted green curve being the exact result, and the solid red and dashed black lines have been perturbed with random noise with different values of $\sigma_\Delta$.

\begin{figure}\centering
\includegraphics[width=0.462\textwidth]{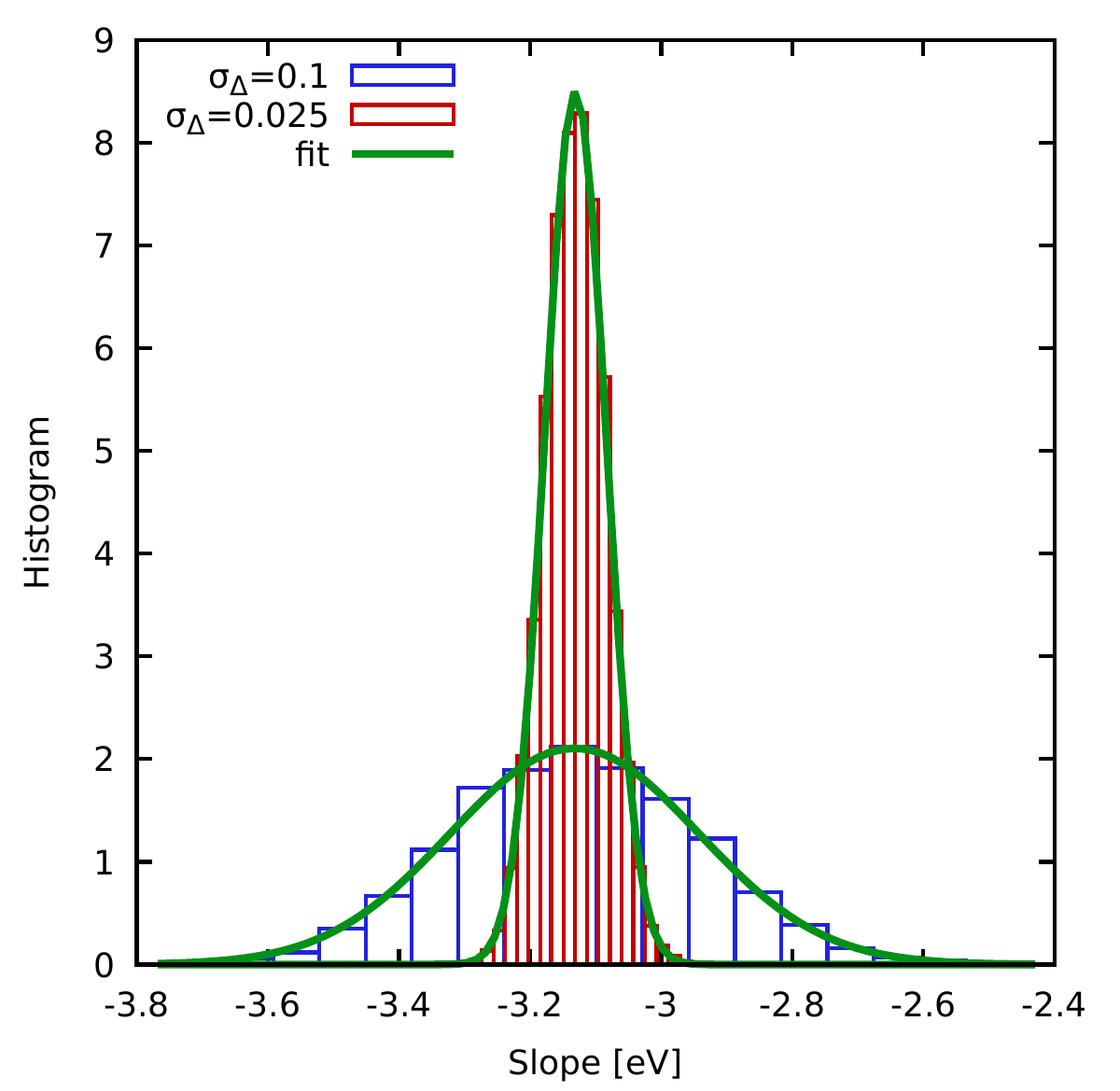}\includegraphics[width=0.462\textwidth]{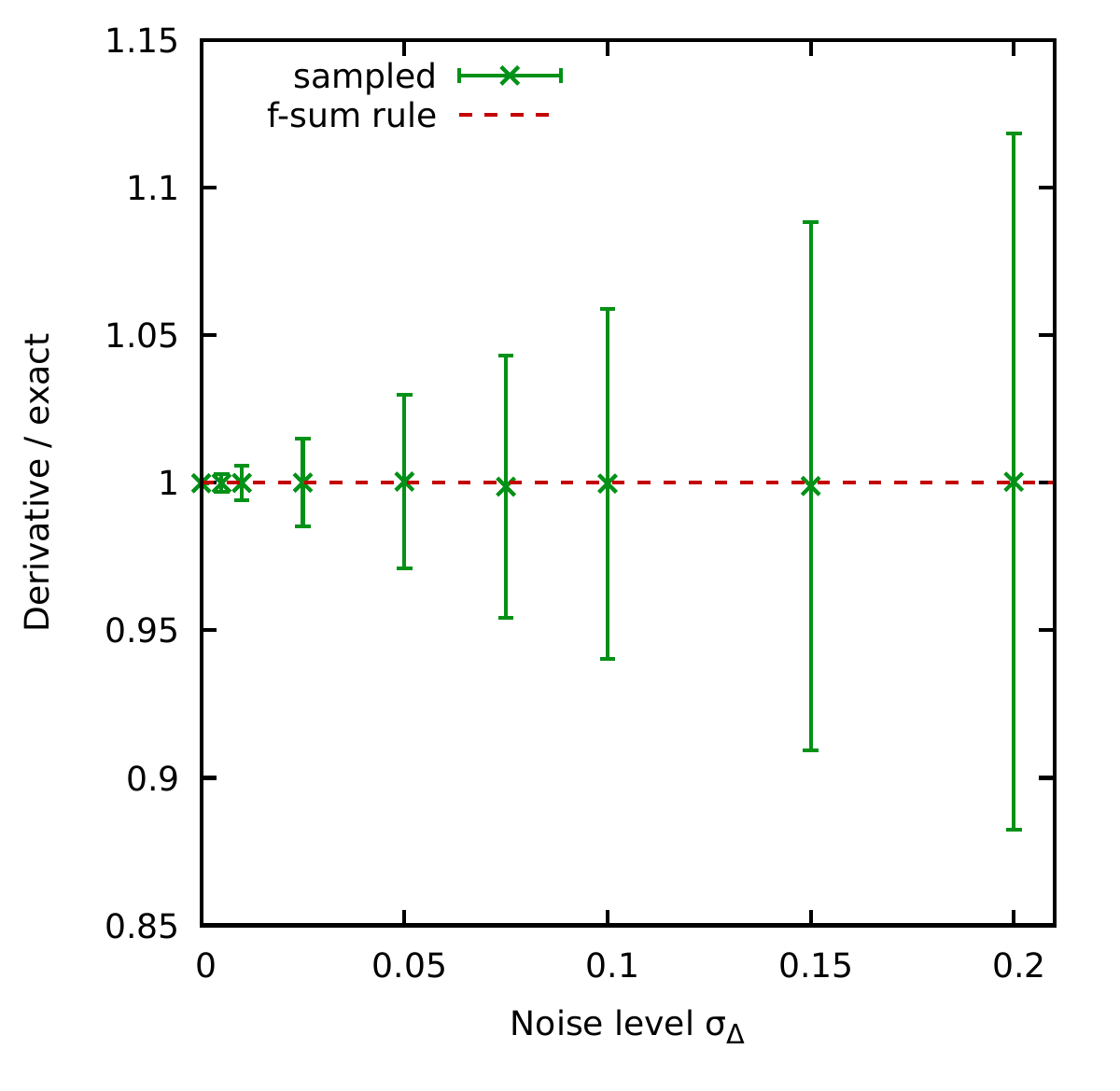}
\caption{\label{fig:synthetic_noise_analysis}
  Sampling $K=10^4$ noisy intensity signals via Eq.~(\ref{eq:I_error}) for each considered value of the noise level $\sigma_\Delta$. a) histogram of computed slopes from the $K$ noisy $I$ samples for $\sigma_\Delta=0.1$ and $\sigma_\Delta=0.025$, cf.~Fig.~\ref{fig:synthetic_noise}. b) estimated variance of slope samples as a function of $\sigma_\Delta$. 
}
\end{figure}

To rigorously quantify the impact of the random noise onto the numerical derivative of the ITCF, we compute $K=10^4$ noisy synthetic intensity signals for different values of $\sigma_\Delta$. In the left panel of Fig.~\ref{fig:synthetic_noise_analysis}, we show corresponding histograms (bars) for $\sigma_\Delta=0.025$ (red) and $\sigma_\Delta=0.1$ (blue). The solid green curves depict Gaussian fits that are in excellent agreement to the histograms. In other words, the presence of random noise governed by Eq.~(\ref{eq:I_error}) leads to a Gaussian distribution of the slope of the corresponding ITCF around the exact value.

This can be seen particularly well in the right panel of Fig.~\ref{fig:synthetic_noise_analysis}, where we show the mean value over $K=10^4$ samples (green crosses) as a function of $\sigma_\Delta$, with the error bars depicting the corresponding variance. Clearly, the mean values nicely reproduce the exact value (dashed red). Moreover, the variance appears to be linear with $\sigma_\Delta$. Indeed, even for the largest considered noise level of $\sigma_\Delta=0.15$, we find a variance within $10\%$ of the exact normalization.

\subsection*{Strongly compressed beryllium}\label{sec:beryllium}

\begin{figure}\centering
\includegraphics[width=0.462\textwidth]{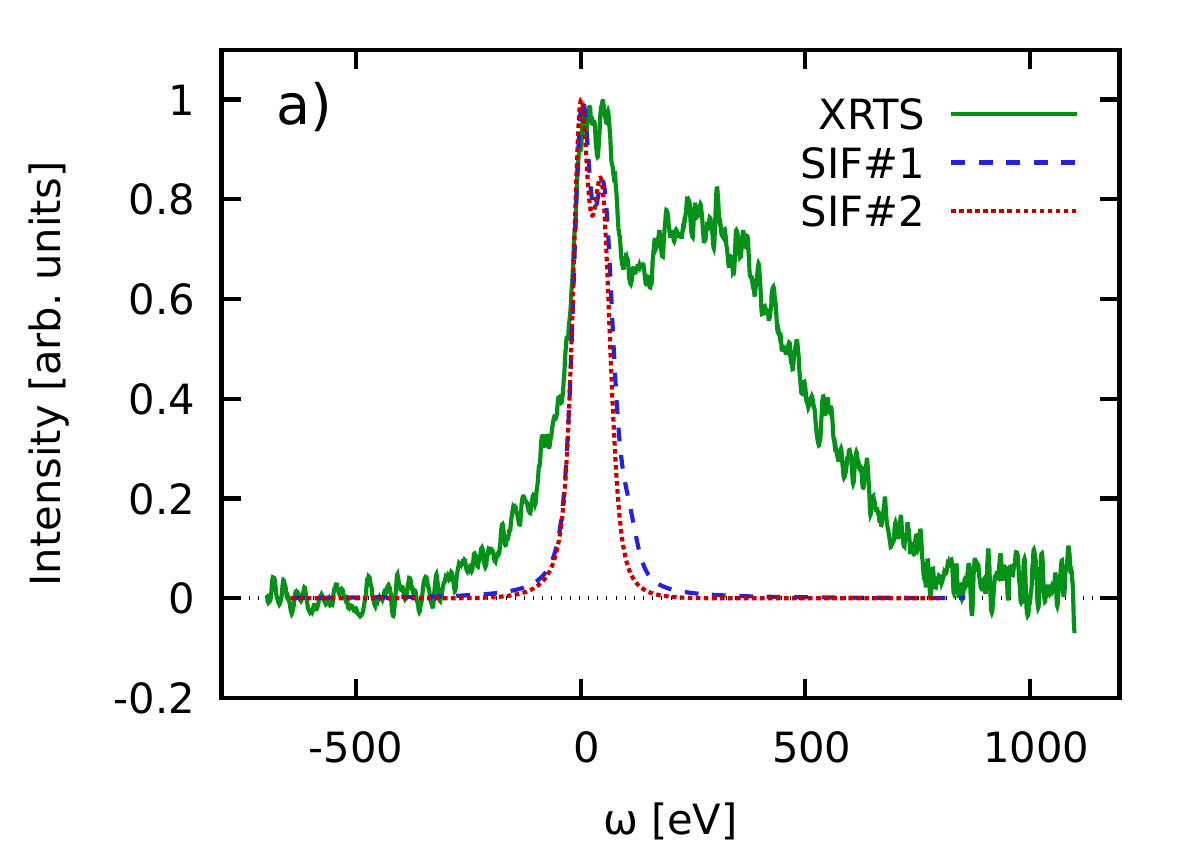}\includegraphics[width=0.462\textwidth]{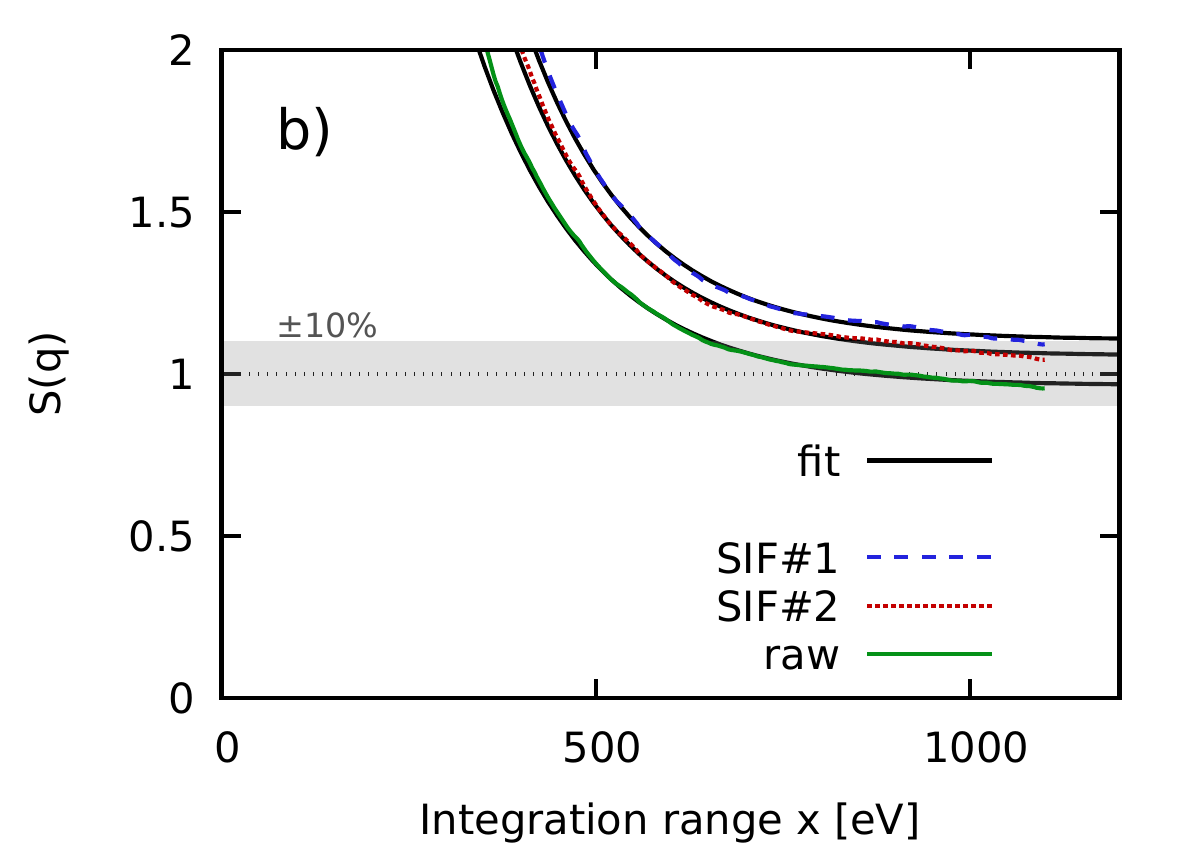}
\caption{\label{fig:NIF}
a) XRTS intensity (solid green) and model SIF (dashed blue) for a NIF shot of strongly compressed Be~\cite{Tilo_Nature_2023} with a nominal temperature and density of $T=119^{+10}_{-50}\,$eV and $n=1.9\pm0.2\times10^{24}\,$cm$^{-3}$. The scattering angle is given by $\theta=120^\circ$, corresponding to $q=7.89\,$\AA$^{-1}$. b) Convergence of the normalized static structure factor $S_{ee}(\mathbf{q})=F_{ee}(\mathbf{q},0)$ as a function of the symmetrically truncated integration range $x$. Dashed blue: full evaluation of Eq.~(\ref{eq:final}); solid green: setting $\mathcal{L}[R(\omega)]\equiv1$; dotted red: truncating the wings of the model SIF at $y=\pm150\,$eV.
}
\end{figure}

In Fig.~\ref{fig:NIF}a), we show an XRTS signal (solid green) that has been obtained for warm dense beryllium at the NIF by D\"oppner \emph{et al.}~\cite{Tilo_Nature_2023}. Capsules with an outer layer of beryllium and a core of air were compressed using 184 laser beams of the NIF and an additional 8 laser beams where used as backlighters to generate the x-rays for the scattering diagnostics. 
The created conditions at $t=0.48$~ns after peak x-ray emission were diagnosed to be at a temperature of $T=119^{+10}_{-50}\,$eV and a density of $n=1.9\pm0.2\times10^{24}\,$cm$^{-3}$ corresponding to five-fold compression of solid beryllium~\cite{Tilo_Nature_2023}. 
In addition, the dashed blue and dotted red curves correspond to two model SIFs taken from the original Ref.~\cite{Tilo_Nature_2023} and the more recent analysis by B\"ohme \emph{et al.}~\cite{boehme2023evidence}, respectively.

A practical obstacle regarding the evaluation of Eq.~(\ref{eq:final}) is given by the limited frequency range of the detector. In practice, we symmetrically truncate the integration range for the two-sided Laplace transform of $I(\mathbf{q},\omega)$ [Eq.~(\ref{eq:Laplace})] at $\pm x$. In Fig.~\ref{fig:NIF}b), we show the convergence of the thus determined, properly normalized static structure factor $S_{ee}(\mathbf{q})=F_{ee}(\mathbf{q},0)$ with $x$. The solid green curve has been computed without taking into account the impact of $R(\omega)$ [i.e., setting $\mathcal{L}[R(\omega)]\equiv 1$ in Eq.~(\ref{eq:final})], whereas the dashed blue and dotted red curves are based on the full evaluation of Eq.~(\ref{eq:final}) using the two different model SIFs.
Empirically, we observe an exponential convergence of $S_{ee}(\mathbf{q})$ with the integration range that is well reproduced by the ansatz
\begin{eqnarray}\label{eq:exponential_fit}
    f(x) = A + B e^{-Cx}\ ,
\end{eqnarray}
where $A$, $B$ and $C$ are free parameters.
The corresponding fits are included as the solid black curves, and are in very good agreement with the input data for $x\gtrsim500\,$eV. The slight dip for $x\gtrsim1000\,$eV is likely an artefact due to a spurious small bump in the measured XRTS signal, see panel a).

From a physical perspective, all three curves
 exhibit the same qualitative behavior and converge around $S_{ee}(\mathbf{q})=1$ in the limit of large $x$. This is indeed the expected vicinity for $S_{ee}(\mathbf{q})$ of Be in the single-particle limit of large $q$ at these conditions, which further substantiates our approach.
A second important point is given by the impact of the SIF model, which has noticeable consequences for the determination of the normalization and, in this way, for the extracted electron--electron static structure factor $S_{ee}(\mathbf{q})$.
This clearly highlights the need for improved source function characterization~\cite{MacDonald_POP_2022} for future experiments.

\begin{figure}\centering
\includegraphics[width=0.462\textwidth]{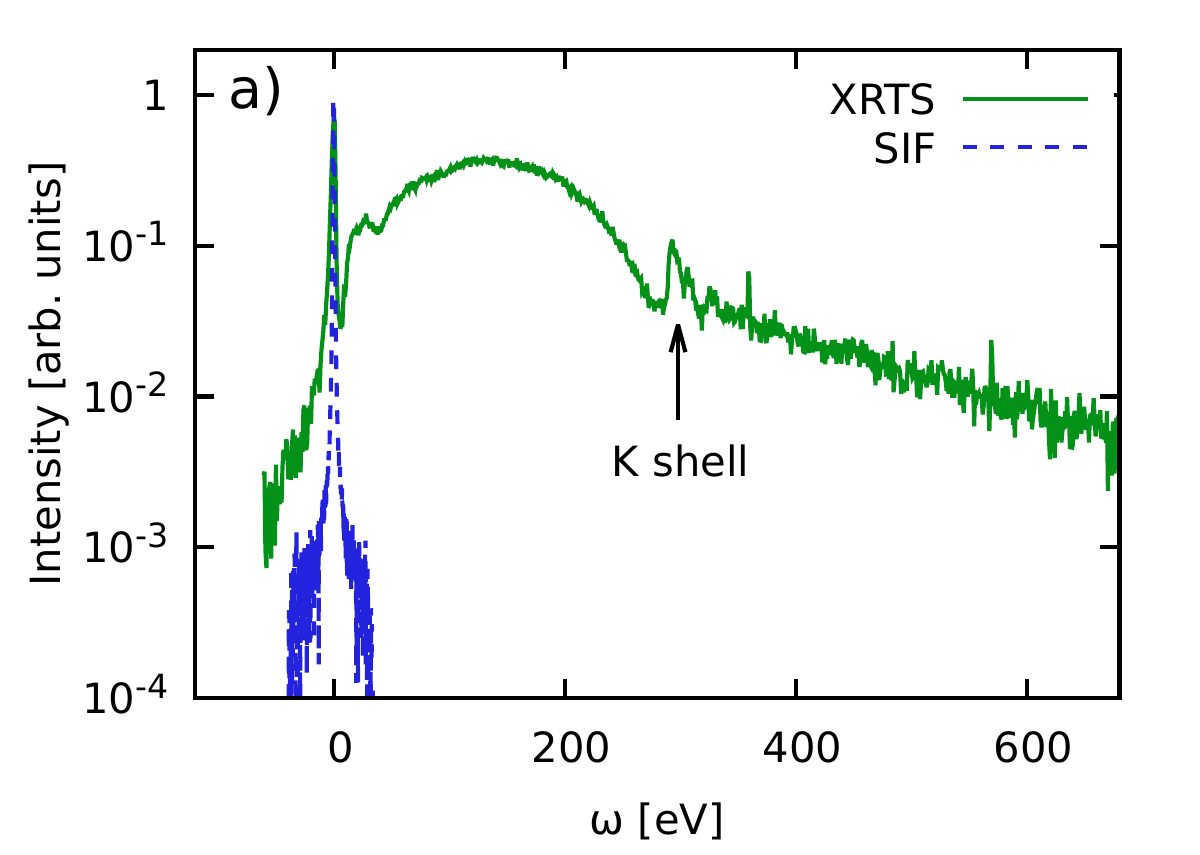}\includegraphics[width=0.462\textwidth]{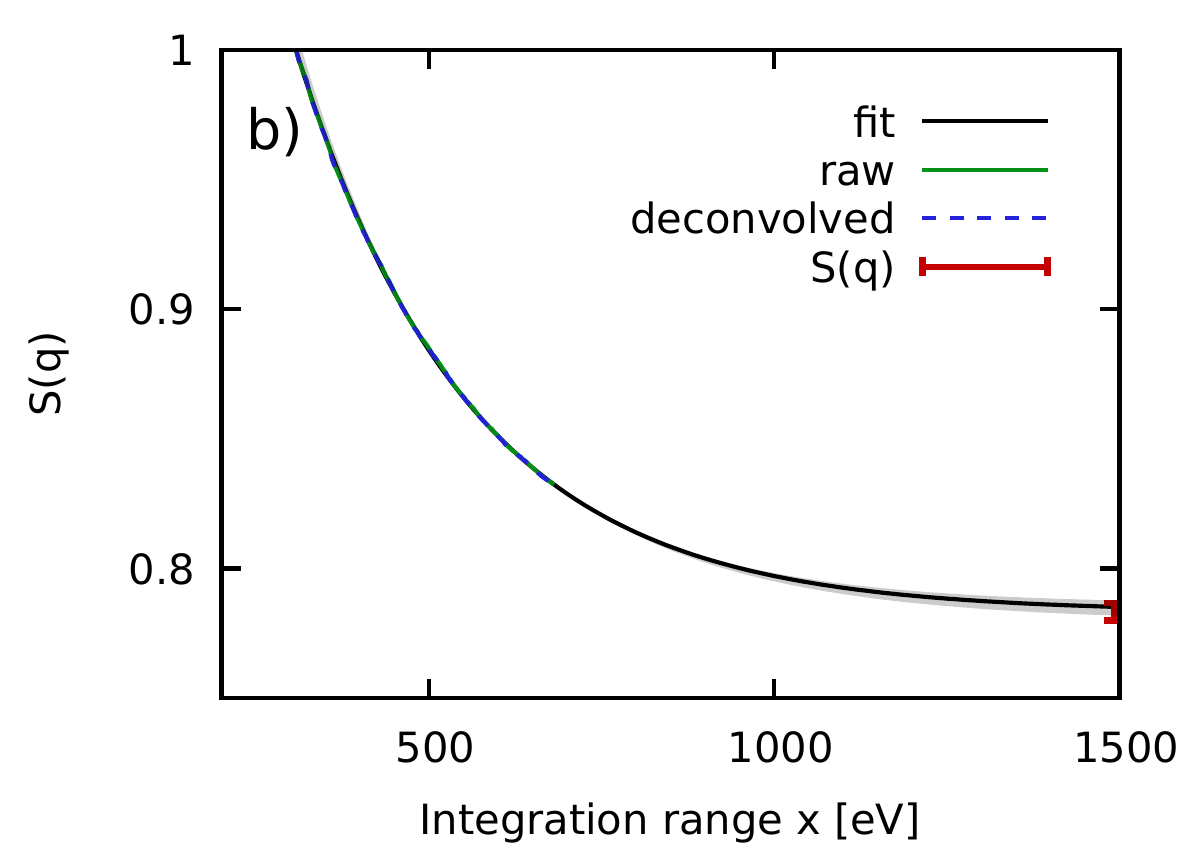}
\caption{\label{fig:carbon} a) XRTS intensity (solid green) and SIF (dashed blue) of carbon measured by
Voigt \emph{et al.}~\cite{Voigt_POP_2021} at the European XFEL. The scattering angle and wave number are given by $\theta=155^\circ$ and $q=5.94\,$\AA$^{-1}$. b) Corresponding convergence of the normalized static structure factor $S_{ee}(\mathbf{q})=F_{ee}(\mathbf{q},0)$ with the integration range $x$. The solid black curve shows an exponential fit to the deconvolved data [cf.~Eq.~(\ref{eq:exponential_fit})], and the red data point marks our final estimation of $S_{ee}(\mathbf{q})=0.783\pm0.003$. 
}
\end{figure}

\subsection*{Diamond at ambient conditions}\label{sec:diamond}

To demonstrate the versatility of our idea, we next consider an XRTS measurement of diamond at ambient conditions that has been performed at the European XFEL by Voigt~\emph{et al.}~\cite{Voigt_POP_2021} in Fig.~\ref{fig:carbon}. The wave number is given by $q=5.94\,$\AA$^{-1}$, which is close to, though not directly on top of, a Bragg peak. This makes the correct value of $S_{ee}(\mathbf{q})$ unclear prior to our analysis. We note the high accuracy of the measured intensity (solid green) over four orders of magnitude, and the remarkably narrow width of the SIF (dashed blue). In addition, Voigt \emph{et al.}~\cite{Voigt_POP_2021} have captured the entire relevant down-shifted (i.e., $\omega>0$) spectral range well beyond the K-shell feature around $\omega=290\,$eV. This is essential for the proper estimation of the normalization as we shall elaborate in more detail in the discussion below.
In Fig.~\ref{fig:carbon}b), we show the corresponding convergence of $S_{ee}(\mathbf{q})$ with $x$. Evidently, we cannot resolve any significant impact of the SIF, which is unsurprising given its narrow bandwidth. We observe a monotonic behavior for $x \gtrsim 400\,$eV that is again well reproduced by the exponential ansatz given by Eq.~(\ref{eq:exponential_fit}) above.

To get an empirical estimate for the associated uncertainty, we have performed multiple fits between $x\geq 300\,$eV and $x\geq 600\,$eV, and the resulting set of possible fits is depicted as the light grey area.
This leads to our final estimate for the static structure factor as $S_{ee}(\mathbf{q})=0.783\pm0.003$. 

\subsection*{Frequency-resolved contributions and implications}\label{sec:resolved}

\begin{figure}[h!]\centering
\includegraphics[width=0.462\textwidth]{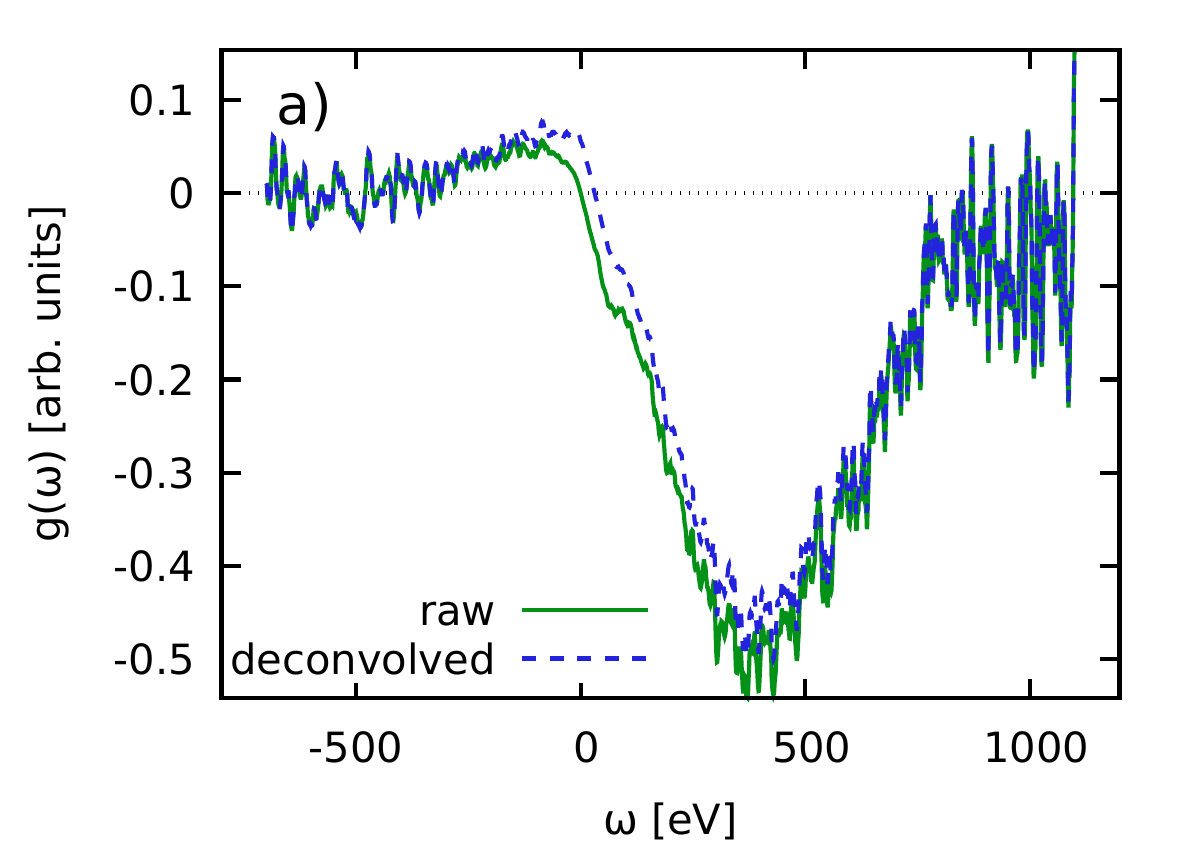}\includegraphics[width=0.462\textwidth]{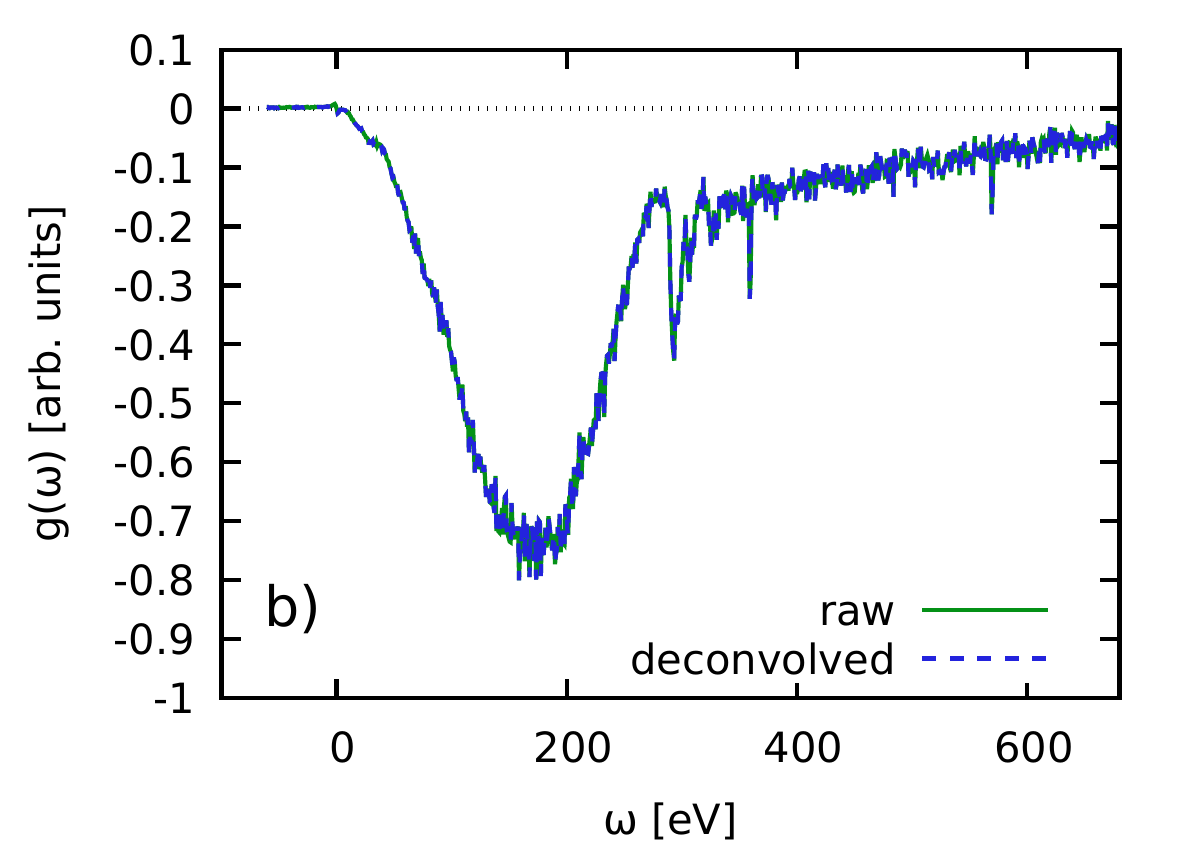}
\caption{\label{fig:NIF_appendix}
Spectrally resolved contribution to the numerical derivative of the ITCF around $\tau=0$, cf.~Eqs.~(\ref{eq:contribution}) and (\ref{eq:epsilon}). Panels a) and b) correspond to the Be NIF shot (Fig.~2 of the main text) and the carbon XFEL shot (Fig.~3 of the main text), respectively, that are analyzed in the main text.
}
\end{figure}

Let us finish our discussion of the imaginary-time evaluation of the f-sum rule by considering 
the spectrally resolved contribution to the numerical derivative Eq.~(\ref{eq:final}); it is given by\begin{eqnarray}\label{eq:contribution}
 g(\omega) = \lim_{\epsilon\to0} f_\epsilon(\omega)\ ,
 \end{eqnarray}
 with the definition
 \begin{eqnarray}\label{eq:epsilon}
    f_\epsilon(\omega)= \frac{I(\mathbf{q},\omega)}{\epsilon}\left(
\frac{e^{-\hbar\omega\epsilon}}{\mathcal{L}\left[R\right](\epsilon)} - \frac{1}{\mathcal{L}\left[R)\right](0)}
    \right)\ .
\end{eqnarray}
Thus, Eq.~(\ref{eq:final}) can be re-written as
 \begin{eqnarray}\label{eq:final_revised}
     A = - \frac{2m_e}{(\hbar q)^2}  \int_{-\infty}^\infty \textnormal{d}\omega\ g(\omega)\ .     
 \end{eqnarray}
In Fig.~\ref{fig:NIF_appendix}a), we show $g(\omega)$ for the Be NIF shot shown in Fig.~\ref{fig:NIF}. Clearly, most contributions to the normalization come from positive scattering energies, $\omega>0$. In fact, the absolute values of the contributions to $A$ qualitatively follow $I(\mathbf{q},\omega)$. This is in stark contrast to the ITCF thermometry method introduced in Refs.~\cite{Dornheim_T_2022,Dornheim_T2_2022}, where negative frequencies contribute with a weight that exponentially increases with $\tau$ due to the exponent in the two-sided Laplace transform, cf.~Eq.~(\ref{eq:Laplace}).
As a consequence, the present approach is available over a broad range of temperatures including cold samples (ambient conditions).
For completeness, we note that the slightly negative contributions around $\omega=1\,$keV are potentially spurious; a possible explanation for this artefact might be the employed background subtraction of the XRTS lineouts, which will be discussed in more detail in a dedicated future publication.

In Fig.~\ref{fig:NIF_appendix}b), we repeat this analysis for the carbon data shown in Fig.~\ref{fig:carbon}. It can be seen that contributions for $\omega<0$ are negligible in practice and we observe a fast, monotonic decrease of $g(\omega)$ with $\omega$.

\section*{Discussion}

 We have presented a new, simulation-free approach for the computation of the a-priori unknown normalization of XRTS measurements. Our approach works for any degree of collectivity, and over a broad range of temperatures including ambient conditions and the HED regime. Furthermore, it is not restricted to thermal equilibrium and, therefore, can readily be applied e.g.~to self-scattering experiments that cannot be described rigorously with existing simulation capabilities. Moreover, we note the high stability of our method with respect to the inevitable noise in the experimental data, and the observed exponential convergence with respect to the integration boundaries.

We are convinced that our scheme provides the basis for a range of interesting future work. Previously, evaluating the Laplace transform of the bare experimental signal allowed one to exploit the symmetry of the ITCF with respect to $\tau$, giving one access to a single parameter, namely the temperature~\cite{Dornheim_T_2022,Dornheim_T2_2022}.
Having absolute knowledge of the ITCF now unlocks the full information about the system in various ways.
For example, the normalization gives us direct access to the static structure factor $S_{ee}(\mathbf{q})$, which is important in its own right. 
This additional bit of information might help to further constrain the forward modelling of XRTS measurements via chemical models~\cite{Gregori_PRE_2003,siegfried_review,kraus_xrts} or \emph{ab initio} simulations~\cite{Schoerner_PRB_2022_2,dynamic2,Ramakrishna_PRB_2021,Moldabekov_PRR_2023}.
The rigorous quantification of the associated uncertainties in the extracted parameters based on Markov chain Monte Carlo (MCMC) calculations~\cite{Kasim_POP_2019} with and without using our new approach is beyond the scope of the present work and will be pursued in dedicated future projects.

In addition, the present scheme makes XRTS a useful tool for the probing of electron--electron correlations on different length scales. The fact that XRTS measurements allow us to check the integration over the available spectral range can also be useful as a rigorous benchmark for X-ray diffraction~\cite{Vorberger_PRE_2015}.
More importantly, the ITCF $F_{ee}(\mathbf{q},\tau)$, by definition, contains the same information as the dynamic structure factor $S_{ee}(\mathbf{q},\omega)$. Experimental results for the properly normalized ITCF will, therefore, allow us to extract a wealth of additional information from XRTS measurements beyond the temperature~\cite{Dornheim_T_2022,Dornheim_T2_2022}, such as the number density $n$ or the charge state $Z$, thereby completing knowledge of an equation-of-state.
Furthermore, it has been suggested that the ITCF also gives one access to quasi-particle excitation energies and even to nontrivial processes such as the roton-type feature that has recently been reported in an interacting electron gas at low density~\cite{Dornheim_Nature_2022,Dornheim_insight_2022,Dornheim_PTR_2022,Hamann_PRR_2023}. In this context, a particularly important relation is given by the imaginary-time version of the fluctuation--dissipation theorem~\cite{Dornheim_insight_2022},
\begin{eqnarray}\label{eq:chi}
    \chi(\mathbf{q},0) = - n \int_0^\beta \textnormal{d}\tau\ F_{ee}(\mathbf{q},\tau)\ ,
\end{eqnarray}
that connects the ITCF with the static limit of the linear density response function, $\chi(\mathbf{q},0)$.
In particular, Eq.~(\ref{eq:chi}) gives one direct access to the static exchange--correlation kernel $K_\textnormal{xc}(\mathbf{q})$, which is a key property in quantum many-body theory~\cite{Dornheim_review, JCP_LR_TDDFT2023, JCP_2021_benchmark}.
In this way, our idea will open up the possibility to rigorously test existing kernels against XRTS measurements at modern XFEL facilities such as the European XFEL~\cite{Tschentscher_2017}.

From a methodological perspective, future tasks include the rigorous quantification of the remaining uncertainty in the normalization due to effects such as $q$-vector blurring or the potential small dependence of $q$ on the scattering frequency $\omega$~\cite{Dornheim_T2_2022}. An additional issue emerges for XRTS measurements of heavier elements, where the excitation of tightly bound-core electrons to the continuum will lead to contributions in $S_{ee}(\mathbf{q},\omega)$ and, thus, $I(\mathbf{q},\omega)$ that are outside of the available detector range. On the one hand, 
an incomplete spectral range prevents the direct determination of the normalization in either the $\omega$- or $\tau$-representation by default. On the other hand, XRTS is an inelastic spectroscopy and the scattering intensity contributed by processes involving core electrons will decay rapidly with increasing binding energy unless the probe beam energy is nearly resonant with a real transition~\cite{humphries2020probing}.
One can imagine quantifying the uncertainty that the finite detection bandwidth introduces by predicting the weight of low-intensity high-energy features associated with bound-free~\cite{souza2014predictions} and bound-bound~\cite{baczewski2021predictions} transitions involving deeply bound core electrons with highly efficient all-electron average atom methods~\cite{johnson2012thomson}.
Further methodological improvements include improved statistical error quantification models, as well as taking into account potential sources of systematic error such as incomplete knowledge of the SIF or the subtraction of a radiation background in the XRTS intensity.



\section*{Acknowledgements}
This work was partially supported by the Center for Advanced Systems Understanding (CASUS) which is financed by Germany’s Federal Ministry of Education and Research (BMBF) and by the Saxon state government out of the State budget approved by the Saxon State Parliament. This work has received funding from the European Research Council (ERC) under the European Union’s Horizon 2022 research and innovation programme
(Grant agreement No. 101076233, "PREXTREME"). 
Views and opinions expressed are however those of the authors only and do not necessarily reflect those of the European Union or the European Research Council Executive Agency. Neither the European Union nor the granting authority can be held responsible for them. The work of Ti.~D.~ and M.~J.~M.~was performed under the auspices of the U.S. Department of Energy by Lawrence Livermore National Laboratory under Contract No. DE-AC52-07NA27344. 
A.D.B.~acknowledges support from Sandia's Laboratory Directed Research and Development Program and US Department of Energy Science Campaign 1. Sandia National Laboratories is a multimission laboratory managed and operated by National Technology and Engineering Solutions of Sandia, LLC, a wholly-owned subsidiary of Honeywell International Inc., for the U.S. Department of Energy’s National Nuclear Security Administration under contract DE-NA0003525.
The PIMC calculations for the synthetic UEG data were carried out at the Norddeutscher Verbund f\"ur Hoch- und H\"ochstleistungsrechnen (HLRN) under grant shp00026, and on a Bull Cluster at the Center for Information Services and High Performance Computing (ZIH) at Technische Universit\"at Dresden.

\section*{Author contributions statement}

To.D.~developed the idea, carried out the analysis, and wrote substantial parts of the ms. Ti.D., A.D.B., P.T., M.P.B., Zh.A.M., Th.G., D.R., D.A.C., M.J.M., Th.R.P., D.K., and J.V., contributed to the analysis, and to writing the ms.




\bibliography{ref}

\begin{thebibliography}{10}
\urlstyle{rm}
\expandafter\ifx\csname url\endcsname\relax
  \def\url#1{\texttt{#1}}\fi
\expandafter\ifx\csname urlprefix\endcsname\relax\def\urlprefix{URL }\fi
\expandafter\ifx\csname doiprefix\endcsname\relax\def\doiprefix{DOI: }\fi
\providecommand{\bibinfo}[2]{#2}
\providecommand{\eprint}[2][]{\url{#2}}

\bibitem{drake2018high}
\bibinfo{author}{Drake, R.}
\newblock \emph{\bibinfo{title}{High-Energy-Density Physics: Foundation of Inertial Fusion and Experimental Astrophysics}}.
\newblock Graduate Texts in Physics (\bibinfo{publisher}{Springer International Publishing}, \bibinfo{year}{2018}).

\bibitem{fortov_review}
\bibinfo{author}{Fortov, V.~E.}
\newblock \bibinfo{journal}{\bibinfo{title}{Extreme states of matter on earth and in space}}.
\newblock {\emph{\JournalTitle{Phys.-Usp}}} \textbf{\bibinfo{volume}{52}}, \bibinfo{pages}{615--647} (\bibinfo{year}{2009}).

\bibitem{Benuzzi_Mounaix_2014}
\bibinfo{author}{Benuzzi-Mounaix, A.} \emph{et~al.}
\newblock \bibinfo{journal}{\bibinfo{title}{Progress in warm dense matter study with applications to planetology}}.
\newblock {\emph{\JournalTitle{Phys. Scripta}}} \textbf{\bibinfo{volume}{T161}}, \bibinfo{pages}{014060}, \doiprefix\url{10.1088/0031-8949/2014/t161/014060} (\bibinfo{year}{2014}).

\bibitem{becker}
\bibinfo{author}{Becker, A.} \emph{et~al.}
\newblock \bibinfo{journal}{\bibinfo{title}{Ab initio equations of state for hydrogen (h-reos.3) and helium (he-reos.3) and their implications for the interior of brown dwarfs}}.
\newblock {\emph{\JournalTitle{Astrophys. J. Suppl. Ser}}} \textbf{\bibinfo{volume}{215}}, \bibinfo{pages}{21} (\bibinfo{year}{2014}).

\bibitem{falk_wdm}
\bibinfo{author}{Falk, K.}
\newblock \bibinfo{journal}{\bibinfo{title}{Experimental methods for warm dense matter research}}.
\newblock {\emph{\JournalTitle{High Power Laser Sci. Eng}}} \textbf{\bibinfo{volume}{6}}, \bibinfo{pages}{e59} (\bibinfo{year}{2018}).

\bibitem{nguyen2022direct}
\bibinfo{author}{Nguyen, Q. L.~D.} \emph{et~al.}
\newblock \bibinfo{journal}{\bibinfo{title}{Direct observation of enhanced electron-phonon coupling in copper nanoparticles in the warm-dense matter regime}}.
\newblock {\emph{\JournalTitle{Phys. Rev. Lett.}}} \textbf{\bibinfo{volume}{131}}, \bibinfo{pages}{085101}, \doiprefix\url{10.1103/PhysRevLett.131.085101} (\bibinfo{year}{2023}).

\bibitem{hu_ICF}
\bibinfo{author}{Hu, S.~X.}, \bibinfo{author}{Militzer, B.}, \bibinfo{author}{Goncharov, V.~N.} \& \bibinfo{author}{Skupsky, S.}
\newblock \bibinfo{journal}{\bibinfo{title}{First-principles equation-of-state table of deuterium for inertial confinement fusion applications}}.
\newblock {\emph{\JournalTitle{Phys. Rev. B}}} \textbf{\bibinfo{volume}{84}}, \bibinfo{pages}{224109} (\bibinfo{year}{2011}).

\bibitem{Moses_NIF}
\bibinfo{author}{Moses, E.~I.}, \bibinfo{author}{Boyd, R.~N.}, \bibinfo{author}{Remington, B.~A.}, \bibinfo{author}{Keane, C.~J.} \& \bibinfo{author}{Al-Ayat, R.}
\newblock \bibinfo{journal}{\bibinfo{title}{The national ignition facility: Ushering in a new age for high energy density science}}.
\newblock {\emph{\JournalTitle{Physics of Plasmas}}} \textbf{\bibinfo{volume}{16}}, \bibinfo{pages}{041006}, \doiprefix\url{10.1063/1.3116505} (\bibinfo{year}{2009}).

\bibitem{kraus_xrts}
\bibinfo{author}{Kraus, D.} \emph{et~al.}
\newblock \bibinfo{journal}{\bibinfo{title}{Characterizing the ionization potential depression in dense carbon plasmas with high-precision spectrally resolved x-ray scattering}}.
\newblock {\emph{\JournalTitle{Plasma Phys. Control Fusion}}} \textbf{\bibinfo{volume}{61}}, \bibinfo{pages}{014015} (\bibinfo{year}{2019}).

\bibitem{Kraus2016}
\bibinfo{author}{Kraus, D.} \emph{et~al.}
\newblock \bibinfo{journal}{\bibinfo{title}{Nanosecond formation of diamond and lonsdaleite by shock compression of graphite}}.
\newblock {\emph{\JournalTitle{Nature Communications}}} \textbf{\bibinfo{volume}{7}}, \bibinfo{pages}{10970}, \doiprefix\url{10.1038/ncomms10970} (\bibinfo{year}{2016}).

\bibitem{Kraus2017}
\bibinfo{author}{Kraus, D.} \emph{et~al.}
\newblock \bibinfo{journal}{\bibinfo{title}{Formation of diamonds in laser-compressed hydrocarbons at planetary interior conditions}}.
\newblock {\emph{\JournalTitle{Nature Astronomy}}} \textbf{\bibinfo{volume}{1}}, \bibinfo{pages}{606--611}, \doiprefix\url{10.1038/s41550-017-0219-9} (\bibinfo{year}{2017}).

\bibitem{PhysRevLett.129.075001}
\bibinfo{author}{Abu-Shawareb, H.} \emph{et~al.}
\newblock \bibinfo{journal}{\bibinfo{title}{Lawson criterion for ignition exceeded in an inertial fusion experiment}}.
\newblock {\emph{\JournalTitle{Phys. Rev. Lett.}}} \textbf{\bibinfo{volume}{129}}, \bibinfo{pages}{075001}, \doiprefix\url{10.1103/PhysRevLett.129.075001} (\bibinfo{year}{2022}).

\bibitem{Zylstra2022}
\bibinfo{author}{Zylstra, A.~B.} \emph{et~al.}
\newblock \bibinfo{journal}{\bibinfo{title}{Burning plasma achieved in inertial fusion}}.
\newblock {\emph{\JournalTitle{Nature}}} \textbf{\bibinfo{volume}{601}}, \bibinfo{pages}{542--548}, \doiprefix\url{10.1038/s41586-021-04281-w} (\bibinfo{year}{2022}).

\bibitem{icf-collab_prl_24}
\bibinfo{author}{Abu-Shawareb, H.} \emph{et~al.}
\newblock \bibinfo{journal}{\bibinfo{title}{Achievement of target gain larger than unity in an inertial fusion experiment}}.
\newblock {\emph{\JournalTitle{Phys. Rev. Lett.}}} \textbf{\bibinfo{volume}{132}}, \bibinfo{pages}{065102}, \doiprefix\url{10.1103/PhysRevLett.132.065102} (\bibinfo{year}{2024}).

\bibitem{sheffield2010plasma}
\bibinfo{author}{Sheffield, J.}, \bibinfo{author}{Froula, D.}, \bibinfo{author}{Glenzer, S.} \& \bibinfo{author}{Luhmann, N.}
\newblock \emph{\bibinfo{title}{Plasma Scattering of Electromagnetic Radiation: Theory and Measurement Techniques}} (\bibinfo{publisher}{Elsevier Science}, \bibinfo{year}{2010}).

\bibitem{siegfried_review}
\bibinfo{author}{Glenzer, S.~H.} \& \bibinfo{author}{Redmer, R.}
\newblock \bibinfo{journal}{\bibinfo{title}{X-ray thomson scattering in high energy density plasmas}}.
\newblock {\emph{\JournalTitle{Rev. Mod. Phys}}} \textbf{\bibinfo{volume}{81}}, \bibinfo{pages}{1625} (\bibinfo{year}{2009}).

\bibitem{Glenzer_PRL_2007}
\bibinfo{author}{Glenzer, S.~H.} \emph{et~al.}
\newblock \bibinfo{journal}{\bibinfo{title}{Observations of plasmons in warm dense matter}}.
\newblock {\emph{\JournalTitle{Phys. Rev. Lett.}}} \textbf{\bibinfo{volume}{98}}, \bibinfo{pages}{065002}, \doiprefix\url{10.1103/PhysRevLett.98.065002} (\bibinfo{year}{2007}).

\bibitem{DOPPNER2009182}
\bibinfo{author}{Döppner, T.} \emph{et~al.}
\newblock \bibinfo{journal}{\bibinfo{title}{Temperature measurement through detailed balance in x-ray thomson scattering}}.
\newblock {\emph{\JournalTitle{High Energy Density Physics}}} \textbf{\bibinfo{volume}{5}}, \bibinfo{pages}{182--186}, \doiprefix\url{https://doi.org/10.1016/j.hedp.2009.05.012} (\bibinfo{year}{2009}).

\bibitem{Abela:77248}
\bibinfo{author}{Abela, R.} \emph{et~al.}
\newblock \emph{\bibinfo{title}{{XFEL}: {T}he {E}uropean {X}-{R}ay {F}ree-{E}lectron {L}aser - {T}echnical {D}esign {R}eport}} (\bibinfo{publisher}{DESY}, \bibinfo{address}{Hamburg}, \bibinfo{year}{2006}).

\bibitem{Dornheim_T2_2022}
\bibinfo{author}{Dornheim, T.} \emph{et~al.}
\newblock \bibinfo{journal}{\bibinfo{title}{{Imaginary-time correlation function thermometry: A new, high-accuracy and model-free temperature analysis technique for x-ray Thomson scattering data}}}.
\newblock {\emph{\JournalTitle{Physics of Plasmas}}} \textbf{\bibinfo{volume}{30}}, \bibinfo{pages}{042707} (\bibinfo{year}{2023}).

\bibitem{MacDonald_POP_2022}
\bibinfo{author}{MacDonald, M.~J.} \emph{et~al.}
\newblock \bibinfo{journal}{\bibinfo{title}{Demonstration of a laser-driven, narrow spectral bandwidth x-ray source for collective x-ray scattering experiments}}.
\newblock {\emph{\JournalTitle{Physics of Plasmas}}} \textbf{\bibinfo{volume}{28}}, \bibinfo{pages}{032708}, \doiprefix\url{10.1063/5.0030958} (\bibinfo{year}{2021}).

\bibitem{Kasim_POP_2019}
\bibinfo{author}{Kasim, M.~F.}, \bibinfo{author}{Galligan, T.~P.}, \bibinfo{author}{Topp-Mugglestone, J.}, \bibinfo{author}{Gregori, G.} \& \bibinfo{author}{Vinko, S.~M.}
\newblock \bibinfo{journal}{\bibinfo{title}{{Inverse problem instabilities in large-scale modeling of matter in extreme conditions}}}.
\newblock {\emph{\JournalTitle{Physics of Plasmas}}} \textbf{\bibinfo{volume}{26}}, \doiprefix\url{10.1063/1.5125979} (\bibinfo{year}{2019}).
\newblock \bibinfo{note}{112706}.

\bibitem{Falk_HEDP_2012}
\bibinfo{author}{Falk, K.} \emph{et~al.}
\newblock \bibinfo{journal}{\bibinfo{title}{Self-consistent measurement of the equation of state of liquid deuterium}}.
\newblock {\emph{\JournalTitle{High Energy Density Physics}}} \textbf{\bibinfo{volume}{8}}, \bibinfo{pages}{76--80}, \doiprefix\url{https://doi.org/10.1016/j.hedp.2011.11.006} (\bibinfo{year}{2012}).

\bibitem{Falk_PRL_2014}
\bibinfo{author}{Falk, K.} \emph{et~al.}
\newblock \bibinfo{journal}{\bibinfo{title}{Equation of state measurements of warm dense carbon using laser-driven shock and release technique}}.
\newblock {\emph{\JournalTitle{Phys. Rev. Lett.}}} \textbf{\bibinfo{volume}{112}}, \bibinfo{pages}{155003}, \doiprefix\url{10.1103/PhysRevLett.112.155003} (\bibinfo{year}{2014}).

\bibitem{Chihara_1987}
\bibinfo{author}{Chihara, J.}
\newblock \bibinfo{journal}{\bibinfo{title}{Difference in x-ray scattering between metallic and non-metallic liquids due to conduction electrons}}.
\newblock {\emph{\JournalTitle{Journal of Physics F: Metal Physics}}} \textbf{\bibinfo{volume}{17}}, \bibinfo{pages}{295--304}, \doiprefix\url{10.1088/0305-4608/17/2/002} (\bibinfo{year}{1987}).

\bibitem{Gregori_PRE_2003}
\bibinfo{author}{Gregori, G.}, \bibinfo{author}{Glenzer, S.~H.}, \bibinfo{author}{Rozmus, W.}, \bibinfo{author}{Lee, R.~W.} \& \bibinfo{author}{Landen, O.~L.}
\newblock \bibinfo{journal}{\bibinfo{title}{Theoretical model of x-ray scattering as a dense matter probe}}.
\newblock {\emph{\JournalTitle{Phys. Rev. E}}} \textbf{\bibinfo{volume}{67}}, \bibinfo{pages}{026412}, \doiprefix\url{10.1103/PhysRevE.67.026412} (\bibinfo{year}{2003}).

\bibitem{Dornheim_T_2022}
\bibinfo{author}{Dornheim, T.} \emph{et~al.}
\newblock \bibinfo{journal}{\bibinfo{title}{Accurate temperature diagnostics for matter under extreme conditions}}.
\newblock {\emph{\JournalTitle{Nature Communications}}} \textbf{\bibinfo{volume}{13}}, \bibinfo{pages}{7911}, \doiprefix\url{10.1038/s41467-022-35578-7} (\bibinfo{year}{2022}).

\bibitem{Dornheim_insight_2022}
\bibinfo{author}{Dornheim, T.}, \bibinfo{author}{Moldabekov, Z.}, \bibinfo{author}{Tolias, P.}, \bibinfo{author}{Böhme, M.} \& \bibinfo{author}{Vorberger, J.}
\newblock \bibinfo{journal}{\bibinfo{title}{Physical insights from imaginary-time density--density correlation functions}}.
\newblock {\emph{\JournalTitle{Matter and Radiation at Extremes}}} \textbf{\bibinfo{volume}{8}}, \bibinfo{pages}{056601}, \doiprefix\url{10.1063/5.0149638} (\bibinfo{year}{2023}).

\bibitem{Dornheim_review}
\bibinfo{author}{Dornheim, T.} \emph{et~al.}
\newblock \bibinfo{journal}{\bibinfo{title}{Electronic density response of warm dense matter}}.
\newblock {\emph{\JournalTitle{Physics of Plasmas}}} \textbf{\bibinfo{volume}{30}}, \bibinfo{pages}{032705}, \doiprefix\url{10.1063/5.0138955} (\bibinfo{year}{2023}).

\bibitem{Moldabekov_JCTC_2023}
\bibinfo{author}{Moldabekov, Z.}, \bibinfo{author}{B{\"o}hme, M.}, \bibinfo{author}{Vorberger, J.}, \bibinfo{author}{Blaschke, D.} \& \bibinfo{author}{Dornheim, T.}
\newblock \bibinfo{journal}{\bibinfo{title}{{Ab Initio Static Exchange--Correlation Kernel across Jacob's Ladder without Functional Derivatives}}}.
\newblock {\emph{\JournalTitle{Journal of Chemical Theory and Computation}}} \textbf{\bibinfo{volume}{19}}, \bibinfo{pages}{1286--1299}, \doiprefix\url{10.1021/acs.jctc.2c01180} (\bibinfo{year}{2023}).

\bibitem{Moldabekov_non_empirical_hybrid}
\bibinfo{author}{Moldabekov, Z.~A.}, \bibinfo{author}{Lokamani, M.}, \bibinfo{author}{Vorberger, J.}, \bibinfo{author}{Cangi, A.} \& \bibinfo{author}{Dornheim, T.}
\newblock \bibinfo{journal}{\bibinfo{title}{{Non-empirical Mixing Coefficient for Hybrid XC Functionals from Analysis of the XC Kernel}}}.
\newblock {\emph{\JournalTitle{The Journal of Physical Chemistry Letters}}} \textbf{\bibinfo{volume}{14}}, \bibinfo{pages}{1326--1333}, \doiprefix\url{10.1021/acs.jpclett.2c03670} (\bibinfo{year}{2023}).

\bibitem{Maximilian_2023}
\bibinfo{author}{Sch\"orner, M.} \emph{et~al.}
\newblock \bibinfo{journal}{\bibinfo{title}{X-ray thomson scattering spectra from density functional theory molecular dynamics simulations based on a modified {C}hihara formula}}.
\newblock {\emph{\JournalTitle{Phys. Rev. E}}} \textbf{\bibinfo{volume}{107}}, \bibinfo{pages}{065207}, \doiprefix\url{10.1103/PhysRevE.107.065207} (\bibinfo{year}{2023}).

\bibitem{JCP_hybrids_2023}
\bibinfo{author}{Moldabekov, Z.~A.}, \bibinfo{author}{Lokamani, M.}, \bibinfo{author}{Vorberger, J.}, \bibinfo{author}{Cangi, A.} \& \bibinfo{author}{Dornheim, T.}
\newblock \bibinfo{journal}{\bibinfo{title}{{Assessing the accuracy of hybrid exchange-correlation functionals for the density response of warm dense electrons}}}.
\newblock {\emph{\JournalTitle{The Journal of Chemical Physics}}} \textbf{\bibinfo{volume}{158}}, \bibinfo{pages}{094105}, \doiprefix\url{10.1063/5.0135729} (\bibinfo{year}{2023}).

\bibitem{Tilo_Nature_2023}
\bibinfo{author}{D{\"o}ppner, T.} \emph{et~al.}
\newblock \bibinfo{journal}{\bibinfo{title}{Observing the onset of pressure-driven k-shell delocalization}}.
\newblock {\emph{\JournalTitle{Nature}}} \doiprefix\url{10.1038/s41586-023-05996-8} (\bibinfo{year}{2023}).

\bibitem{Voigt_POP_2021}
\bibinfo{author}{Voigt, K.} \emph{et~al.}
\newblock \bibinfo{journal}{\bibinfo{title}{{Demonstration of an x-ray Raman spectroscopy setup to study warm dense carbon at the high energy density instrument of European XFEL}}}.
\newblock {\emph{\JournalTitle{Physics of Plasmas}}} \textbf{\bibinfo{volume}{28}}, \bibinfo{pages}{082701}, \doiprefix\url{10.1063/5.0048150} (\bibinfo{year}{2021}).

\bibitem{Vorberger_PRX_2023}
\bibinfo{author}{Vorberger, J.} \emph{et~al.}
\newblock \bibinfo{journal}{\bibinfo{title}{Revealing non-equilibrium and relaxation in laser heated matter}}.
\newblock {\emph{\JournalTitle{Physics Letters A}}} \textbf{\bibinfo{volume}{499}}, \bibinfo{pages}{129362}, \doiprefix\url{https://doi.org/10.1016/j.physleta.2024.129362} (\bibinfo{year}{2024}).

\bibitem{Dornheim_moments_2023}
\bibinfo{author}{Dornheim, T.} \emph{et~al.}
\newblock \bibinfo{journal}{\bibinfo{title}{Extraction of the frequency moments of spectral densities from imaginary-time correlation function data}}.
\newblock {\emph{\JournalTitle{Phys. Rev. B}}} \textbf{\bibinfo{volume}{107}}, \bibinfo{pages}{155148}, \doiprefix\url{10.1103/PhysRevB.107.155148} (\bibinfo{year}{2023}).

\bibitem{quantum_theory}
\bibinfo{author}{Giuliani, G.} \& \bibinfo{author}{Vignale, G.}
\newblock \emph{\bibinfo{title}{Quantum Theory of the Electron Liquid}} (\bibinfo{publisher}{Cambridge University Press}, \bibinfo{address}{Cambridge}, \bibinfo{year}{2008}).

\bibitem{Dornheim_PTR_2022}
\bibinfo{author}{Dornheim, T.}, \bibinfo{author}{Vorberger, J.}, \bibinfo{author}{Moldabekov, Z.} \& \bibinfo{author}{Böhme, M.}
\newblock \bibinfo{journal}{\bibinfo{title}{Analysing the dynamic structure of warm dense matter in the imaginary-time domain: theoretical models and simulations}}.
\newblock {\emph{\JournalTitle{Phil. Trans. R. Soc. A}}} \textbf{\bibinfo{volume}{381}}, \bibinfo{pages}{20220217}, \doiprefix\url{10.1098/rsta.2022.0217} (\bibinfo{year}{2023}).

\bibitem{dornheim_dynamic}
\bibinfo{author}{Dornheim, T.}, \bibinfo{author}{Groth, S.}, \bibinfo{author}{Vorberger, J.} \& \bibinfo{author}{Bonitz, M.}
\newblock \bibinfo{journal}{\bibinfo{title}{Ab initio path integral {M}onte {C}arlo results for the dynamic structure factor of correlated electrons: From the electron liquid to warm dense matter}}.
\newblock {\emph{\JournalTitle{Phys. Rev. Lett.}}} \textbf{\bibinfo{volume}{121}}, \bibinfo{pages}{255001} (\bibinfo{year}{2018}).

\bibitem{Dornheim_PRL_2020_ESA}
\bibinfo{author}{Dornheim, T.} \emph{et~al.}
\newblock \bibinfo{journal}{\bibinfo{title}{Effective static approximation: A fast and reliable tool for warm-dense matter theory}}.
\newblock {\emph{\JournalTitle{Phys. Rev. Lett.}}} \textbf{\bibinfo{volume}{125}}, \bibinfo{pages}{235001}, \doiprefix\url{10.1103/PhysRevLett.125.235001} (\bibinfo{year}{2020}).

\bibitem{GarciaSaiz2008}
\bibinfo{author}{Garc{\'i}a~Saiz, E.} \emph{et~al.}
\newblock \bibinfo{journal}{\bibinfo{title}{Probing warm dense lithium by inelastic x-ray scattering}}.
\newblock {\emph{\JournalTitle{Nature Physics}}} \textbf{\bibinfo{volume}{4}}, \bibinfo{pages}{940--944}, \doiprefix\url{10.1038/nphys1103} (\bibinfo{year}{2008}).

\bibitem{boehme2023evidence}
\bibinfo{author}{Böhme, M.~P.} \emph{et~al.}
\newblock \bibinfo{journal}{\bibinfo{title}{Evidence of free-bound transitions in warm dense matter and their impact on equation-of-state measurements}}.
\newblock {\emph{\JournalTitle{arXiv}}}  (\bibinfo{year}{2023}).
\newblock \eprint{2306.17653}.

\bibitem{Schoerner_PRB_2022_2}
\bibinfo{author}{Sch\"orner, M.}, \bibinfo{author}{Witte, B. B.~L.}, \bibinfo{author}{Baczewski, A.~D.}, \bibinfo{author}{Cangi, A.} \& \bibinfo{author}{Redmer, R.}
\newblock \bibinfo{journal}{\bibinfo{title}{Ab initio study of shock-compressed copper}}.
\newblock {\emph{\JournalTitle{Phys. Rev. B}}} \textbf{\bibinfo{volume}{106}}, \bibinfo{pages}{054304}, \doiprefix\url{10.1103/PhysRevB.106.054304} (\bibinfo{year}{2022}).

\bibitem{dynamic2}
\bibinfo{author}{Baczewski, A.~D.}, \bibinfo{author}{Shulenburger, L.}, \bibinfo{author}{Desjarlais, M.~P.}, \bibinfo{author}{Hansen, S.~B.} \& \bibinfo{author}{Magyar, R.~J.}
\newblock \bibinfo{journal}{\bibinfo{title}{{X-ray Thomson Scattering in Warm Dense Matter without the Chihara Decomposition}}}.
\newblock {\emph{\JournalTitle{Phys. Rev. Lett}}} \textbf{\bibinfo{volume}{116}}, \bibinfo{pages}{115004} (\bibinfo{year}{2016}).

\bibitem{Ramakrishna_PRB_2021}
\bibinfo{author}{Ramakrishna, K.}, \bibinfo{author}{Cangi, A.}, \bibinfo{author}{Dornheim, T.}, \bibinfo{author}{Baczewski, A.} \& \bibinfo{author}{Vorberger, J.}
\newblock \bibinfo{journal}{\bibinfo{title}{First-principles modeling of plasmons in aluminum under ambient and extreme conditions}}.
\newblock {\emph{\JournalTitle{Phys. Rev. B}}} \textbf{\bibinfo{volume}{103}}, \bibinfo{pages}{125118}, \doiprefix\url{10.1103/PhysRevB.103.125118} (\bibinfo{year}{2021}).

\bibitem{Moldabekov_PRR_2023}
\bibinfo{author}{Moldabekov, Z.~A.}, \bibinfo{author}{Pavanello, M.}, \bibinfo{author}{B\"ohme, M.~P.}, \bibinfo{author}{Vorberger, J.} \& \bibinfo{author}{Dornheim, T.}
\newblock \bibinfo{journal}{\bibinfo{title}{Linear-response time-dependent density functional theory approach to warm dense matter with adiabatic exchange-correlation kernels}}.
\newblock {\emph{\JournalTitle{Phys. Rev. Res.}}} \textbf{\bibinfo{volume}{5}}, \bibinfo{pages}{023089}, \doiprefix\url{10.1103/PhysRevResearch.5.023089} (\bibinfo{year}{2023}).

\bibitem{Vorberger_PRE_2015}
\bibinfo{author}{Vorberger, J.} \& \bibinfo{author}{Gericke, D.~O.}
\newblock \bibinfo{journal}{\bibinfo{title}{Ab initio approach to model x-ray diffraction in warm dense matter}}.
\newblock {\emph{\JournalTitle{Phys. Rev. E}}} \textbf{\bibinfo{volume}{91}}, \bibinfo{pages}{033112}, \doiprefix\url{10.1103/PhysRevE.91.033112} (\bibinfo{year}{2015}).

\bibitem{Dornheim_Nature_2022}
\bibinfo{author}{Dornheim, T.}, \bibinfo{author}{Moldabekov, Z.}, \bibinfo{author}{Vorberger, J.}, \bibinfo{author}{K{\"a}hlert, H.} \& \bibinfo{author}{Bonitz, M.}
\newblock \bibinfo{journal}{\bibinfo{title}{Electronic pair alignment and roton feature in the warm dense electron gas}}.
\newblock {\emph{\JournalTitle{Communications Physics}}} \textbf{\bibinfo{volume}{5}}, \bibinfo{pages}{304}, \doiprefix\url{10.1038/s42005-022-01078-9} (\bibinfo{year}{2022}).

\bibitem{Hamann_PRR_2023}
\bibinfo{author}{Hamann, P.} \emph{et~al.}
\newblock \bibinfo{journal}{\bibinfo{title}{Prediction of a roton-type feature in warm dense hydrogen}}.
\newblock {\emph{\JournalTitle{Phys. Rev. Res.}}} \textbf{\bibinfo{volume}{5}}, \bibinfo{pages}{033039}, \doiprefix\url{10.1103/PhysRevResearch.5.033039} (\bibinfo{year}{2023}).

\bibitem{JCP_LR_TDDFT2023}
\bibinfo{author}{Moldabekov, Z.~A.}, \bibinfo{author}{Vorberger, J.}, \bibinfo{author}{Lokamani, M.} \& \bibinfo{author}{Dornheim, T.}
\newblock \bibinfo{journal}{\bibinfo{title}{{Averaging over atom snapshots in linear-response TDDFT of disordered systems: A case study of warm dense hydrogen}}}.
\newblock {\emph{\JournalTitle{The Journal of Chemical Physics}}} \textbf{\bibinfo{volume}{159}}, \bibinfo{pages}{014107}, \doiprefix\url{10.1063/5.0152126} (\bibinfo{year}{2023}).

\bibitem{JCP_2021_benchmark}
\bibinfo{author}{Moldabekov, Z.}, \bibinfo{author}{Dornheim, T.}, \bibinfo{author}{Böhme, M.}, \bibinfo{author}{Vorberger, J.} \& \bibinfo{author}{Cangi, A.}
\newblock \bibinfo{journal}{\bibinfo{title}{{The relevance of electronic perturbations in the warm dense electron gas}}}.
\newblock {\emph{\JournalTitle{The Journal of Chemical Physics}}} \textbf{\bibinfo{volume}{155}}, \bibinfo{pages}{124116}, \doiprefix\url{10.1063/5.0062325} (\bibinfo{year}{2021}).

\bibitem{Tschentscher_2017}
\bibinfo{author}{Tschentscher, T.} \emph{et~al.}
\newblock \bibinfo{journal}{\bibinfo{title}{{Photon Beam Transport and Scientific Instruments at the European XFEL}}}.
\newblock {\emph{\JournalTitle{Applied Sciences}}} \textbf{\bibinfo{volume}{7}}, \bibinfo{pages}{592} (\bibinfo{year}{2017}).

\bibitem{humphries2020probing}
\bibinfo{author}{Humphries, O.} \emph{et~al.}
\newblock \bibinfo{journal}{\bibinfo{title}{Probing the electronic structure of warm dense nickel via resonant inelastic x-ray scattering}}.
\newblock {\emph{\JournalTitle{Physical Review Letters}}} \textbf{\bibinfo{volume}{125}}, \bibinfo{pages}{195001}, \doiprefix\url{10.1103/PhysRevLett.125.195001} (\bibinfo{year}{2020}).

\bibitem{souza2014predictions}
\bibinfo{author}{Souza, A.}, \bibinfo{author}{Perkins, D.}, \bibinfo{author}{Starrett, C.}, \bibinfo{author}{Saumon, D.} \& \bibinfo{author}{Hansen, S.}
\newblock \bibinfo{journal}{\bibinfo{title}{Predictions of x-ray scattering spectra for warm dense matter}}.
\newblock {\emph{\JournalTitle{Physical Review E}}} \textbf{\bibinfo{volume}{89}}, \bibinfo{pages}{023108}, \doiprefix\url{10.1103/PhysRevE.89.023108} (\bibinfo{year}{2014}).

\bibitem{baczewski2021predictions}
\bibinfo{author}{Baczewski, A.~D.}, \bibinfo{author}{Hentschel, T.}, \bibinfo{author}{Kononov, A.} \& \bibinfo{author}{Hansen, S.~B.}
\newblock \bibinfo{journal}{\bibinfo{title}{{Predictions of bound-bound transition signatures in x-ray Thomson scattering}}}.
\newblock {\emph{\JournalTitle{arXiv preprint arXiv:2109.09576}}} \doiprefix\url{10.48550/arXiv.2109.09576} (\bibinfo{year}{2021}).

\bibitem{johnson2012thomson}
\bibinfo{author}{Johnson, W.}, \bibinfo{author}{Nilsen, J.}, \bibinfo{author}{Cheng, K.} \emph{et~al.}
\newblock \bibinfo{journal}{\bibinfo{title}{Thomson scattering in the average-atom approximation}}.
\newblock {\emph{\JournalTitle{Physical Review E}}} \textbf{\bibinfo{volume}{86}}, \bibinfo{pages}{036410}, \doiprefix\url{10.1103/PhysRevE.86.036410} (\bibinfo{year}{2012}).

\end{thebibliography}

\end{document}